\documentclass[aps,prc,groupedaddress,superscriptaddress,amsmath,floatfix,twocolumn]{revtex4}

\usepackage{nicefrac}
\usepackage{pst-all}
\usepackage{graphicx}
\definecolor{darkgreen}{rgb}{0,.5,0}
\definecolor{violett}{rgb}{.5,0,.5}
\definecolor{orange}{rgb}{.8,.4,0}

\begin{document}
\title{Direct photon calculations in heavy-ion collisions at $\sqrt{s_{\rm
NN}} = 62.4 - 200$~AGeV in a (3+1) dimensional hybrid approach}
\author{Bj\o{}rn B\"auchle}
\email{baeuchle@th.physik.uni-frankfurt.de}
\author{Marcus Bleicher}
\affiliation{Frankfurt Institute for Advanced Studies, Frankfurt am Main, Germany}
\affiliation{Institut f\"ur Theoretische Physik, Goethe-Universit\"at,
Frankfurt am Main, Germany}

\begin{abstract}

Direct photon spectra from central Au+Au- and Cu+Cu-collisions at
$\sqrt{s_{\rm NN}} = 62.4, 130$ and $200~$AGeV are calculated within the
microscopic transport model UrQMD and a micro+macro hybrid model. In the
latter approach, the high-density part of the transport evolution is
replaced by an ideal 3+1-dimensional hydrodynamic calculation. We study the
impact of viscosity and full local thermalization and compare the
calculations to measurements obtained by the PHENIX collaboration. We find a
reasonable agreement with the experimental data for calculations involving a
Quark-Gluon-Plasma phase.

\end{abstract}

\maketitle

\section{Introduction}\label{sec:intro}

Heavy-ion physics is widely used as a tool for the exploration of the phase
diagram of strongly interacting matter. In the collision of heavy nuclei,
the nucleons may be compressed and heated sufficiently to create a new state of
matter that consists of partonic degrees of freedom, the Quark-Gluon-Plasma
(QGP)~\cite{Harris:1996zx,Bass:1998vz}. Indeed, proposed signatures for the QGP, like
strong jet quenching and large elliptic flow have been found by experiments
at the Relativistic Heavy Ion Collider
(BNL-RHIC)~\cite{nucl-ex/0304022,nucl-ex/0308006,Adams:2005dq,Back:2004je,Arsene:2004fa,Adcox:2004mh}.

Inferring knowledge about the central regions of a heavy-ion collision is
very difficult, since even if a plasma is created, its lifetime and size are
beyond the experimental reach for direct observation, so we are limited to
the study of particles that are emitted from the reaction zone.
Unfortunately, first principle calculations of QCD-processes are only
possible if all involved scales are much larger than the QCD-scale
$\Lambda_{\rm QCD} \approx 0.2$~GeV.  However, in a heavy-ion collision,
most particles have momenta comparable to $\Lambda_{\rm QCD}$. Therefore,
more phenomenological approaches are necessary to explore the bulk of the
matter.

While the abundance of hadronic particles that are produced in a heavy-ion collision are
emitted at the end of the reaction and carry only indirect information from
the early stages, electromagnetic probes allow for an undisturbed view into
all stages of the reaction. Photons and leptons escape the reaction zone
without rescattering due to their
very small cross-section, but for the same reason, their abundancies are rather
low, compared to hadronic species~\cite{arXiv:0904.2184}.

Three different electromagnetic particle species are currently being
measured in heavy-ion experiments: single- and dielectrons, single- and
dimuons and photons. Direct photons have the advantage that they are created
in scatterings of the partonic or hadronic medium and are therefore directly
coupled to the region of interaction. The leptons, however, are usually
created in pairs, either in the (initial state) Drell Yan process or by the
decay of hadrons. In addition, one of the leptons might be a neutrino, which
escapes observation.  Since this process is governed by the weak
interaction, the decay usually happens outside the fireball. Single leptons
are therefore used to reconstruct weakly decaying heavy quarks, while the
invariant mass distribution of dileptons can be used to extract spectral
functions of vector mesons.

Previous calculations of direct photons from transport theory include work
with UrQMD by Dumitru {\it et al.}~\cite{hep-ph/9709487} and B\"auchle {\it
et al.}~\cite{arXiv:0905.4678} and with HSD by Bratkovskaya {\it et
al.}~\cite{arXiv:0806.3465}. Hydrodynamics has been used in many direct
photon calculations, see
e.g.~\cite{Kapusta:1991qp,nucl-th/0006018,Turbide:2003si,hep-ph/0502248,arXiv:0811.0666,arXiv:0902.1303,arXiv:0903.1764,arXiv:0911.2426}.

The extraction of the yield of photons from the fireball (direct
photons) is hindered by a huge background of photons from hadronic decays
outside the fireball, which is dominated by the $\pi^0$- and $\eta$-decays.
However, experimental techniques for the extraction of direct photon yields
are well developed and allow to disentangle these late stage contributions
from the scattering contribution. The experimental methods include a
direct estimation of the background via invariant mass-analysis of the
photons~\cite{nucl-ex/0006007,nucl-ex/0006008}, the analysis of interference
patterns (using a Hanburry Brown-Twiss analysis)~\cite{nucl-ex/0310022} and
the extrapolation of the spectra of low-mass dileptons to the photon
point~\cite{arXiv:0912.0244}.

In this paper, we apply a previously established model for direct photon
emission from hadronic and partonic sources~\cite{arXiv:0905.4678} and
apply it to collision systems measured by the STAR and PHENIX collaborations
at BNL-RHIC. In Section~\ref{sec:hybrid}, we briefly introduce the model and
the parameters used for the present calculations, and in Section~\ref{sec:results}
we show the direct photon spectra obtained with our calculations as well as
comparisons to the available data from the PHENIX
collaboration~\cite{nucl-ex/0503003,arXiv:0804.4168}.

\section{The model}\label{sec:hybrid}

In the present work, direct photon spectra are calculated in the framework
of the microscopic Ultrarelativistic Quantum Molecular Dynamics (UrQMD)
transport model~\cite{Bass:1998ca,Bleicher:1999xi,Petersen:2008kb}, using
the hybrid option introduced in version
3.3~\cite{Rischke:1995ir,Rischke:1995mt,Petersen:2008dd,u3.3}.  While UrQMD
itself is a hadronic transport model that includes only hadronic and string
degrees of freedom and employs PYTHIA~\cite{Sjostrand:2006za} for
scatterings at high momentum transfer, the hybrid option allows to
substitute the high-density part of the evolution by a 3+1-dimensional ideal
hydrodynamic~\cite{Rischke:1995mt} description. In this part,
other-than-hadronic degrees of freedom and phase transitions may be
included.

The inclusion of an intermediate phase into the model raises the need for
two interfaces, to go from the particle-based description of the transport
model to the density-based description of the hydrodynamic model and back
again.

The mapping from transport simulation to hydrodynamics is performed at
$t_{\rm start} = 0.6$~fm. Here, the energy-density, baryon number-density
and momentum densities are calculated from all particles at midrapidity.
Particles with a rapidity $|y| > 2$ are propagated in the cascade and do not
interact with the bulk medium.

The transition from hydrodynamics back to the cascade proceeds gradually,
mapping the temperatures and chemical potentials to particles via the
Cooper-Frye-formula~\cite{Cooper:1974mv} when all cells in the same
transverse slice (i.e.\ at the same position along the beam direction) have
diluted below a critical energy density (see Table~\ref{tab:epscrit}).
After the transition to the cascade, rescatterings and decays are calculated
in the well-known UrQMD model. For more detailed information on the hybrid
model the reader is referred to~\cite{Petersen:2008dd,arXiv:0905.3099}. 

\subsection{Equations of State}\label{sec:hybrid:eos}

Three different Equations of State (EoS) are compared in this work. The
effects of thermalization at the transition from the initial stage cascade
to hydrodynamics can be explored with the Hadron Gas-EoS
(HG-EoS)~\cite{nucl-th/0209022}, which has the same degrees of freedom as
the transport phase. To investigate the effects of partonic matter and a
phase transition, we use two different models for the EoS: The Chiral
Equation of State $\chi$-EoS~\cite{arXiv:0909.4421} has a cross-over phase
transition to chirally restored and deconfined matter, while the Bag Model
Equation of State BM-EoS~\cite{Rischke:1995mt} has a first order phase
transition to a Quark Gluon Plasma. In both EoS, the transition happens at
around $T_{\rm C} \approx 170$~MeV.

\begin{table}
 \begin{tabular}{l|r}
  EoS & $\epsilon_{\rm crit}$ \\ \hline
  HG-EoS  & $5 \epsilon_0$ \\
  $\chi$-EoS & $7 \epsilon_0$ \\
  BM-EoS & $5 \epsilon_0$
 \end{tabular}
 \caption{The critical energy densities for the mapping from hydrodynamics
 to transport theory for the various Equations of state. $\epsilon_0 =
 146$~MeV/fm$^3$ is the nuclear ground state energy density.}
 \label{tab:epscrit}
\end{table}

\subsection{Photon emission sources}\label{sec:photons}

Due to the small creation probability of direct photons, their emission is
calculated perturbatively. I.e., the evolution of the underlying event
remains unaltered by the emission of direct photons.

The set of channels for direct photon production differ in the transport and
hydrodynamic parts of the model. The most important channels, though, are
common to both parts, namely $\pi\pi\rightarrow\gamma\rho$ and
$\pi\rho\rightarrow\gamma\pi$. Besides photon emission from the
Quark-Gluon-Plasma, channels with strangeness are included in the
hydrodynamic part. The corresponding rates for photon emission from each
hydrodynamic cell are taken from Turbide {\it et al.}~\cite{Turbide:2003si}.
In the transport part, additional processes including an $\eta$-meson are
included. The corresponding cross-sections have been calculated by Kapusta
{\it et al.}~\cite{Kapusta:1991qp}.

Although Kapusta and Turbide use different Lagrangians to derive their
cross-sections and rates, earlier investigations
(see~\cite{arXiv:0905.4678}) have shown that the thermal rates that can be
extracted from Kapusta's cross-sections using this model agree very well
with those parametrized by Turbide {\it et al.}. The same investigations
have shown that the contributions of the hadronic processes that are not
common to both models contribute about equally, but not significantly to the
final spectra.  The numerical implementation for direct photon emission is
explained in detail in~\cite{arXiv:0905.4678}.

At high transverse momenta, another source becomes important, namely the
prompt contribution from hard scatterings of partons in the initial nuclei.
The spectra predicted by NLO-pQCD calculations from Gordon and
Vogelsang~\cite{Gordon:1993qc} fit the experimental data from the
PHENIX-collaboration~\cite{nucl-ex/0503003} rather well at high $p_\bot$.
Therefore, the pQCD contributions from~\cite{Gordon:1993qc}, scaled by the
number of binary collisions $\langle N_{\sf coll} \rangle$ are added to the
soft photons calculated here.

\section{Results}\label{sec:results}

\begin{figure}
 \input{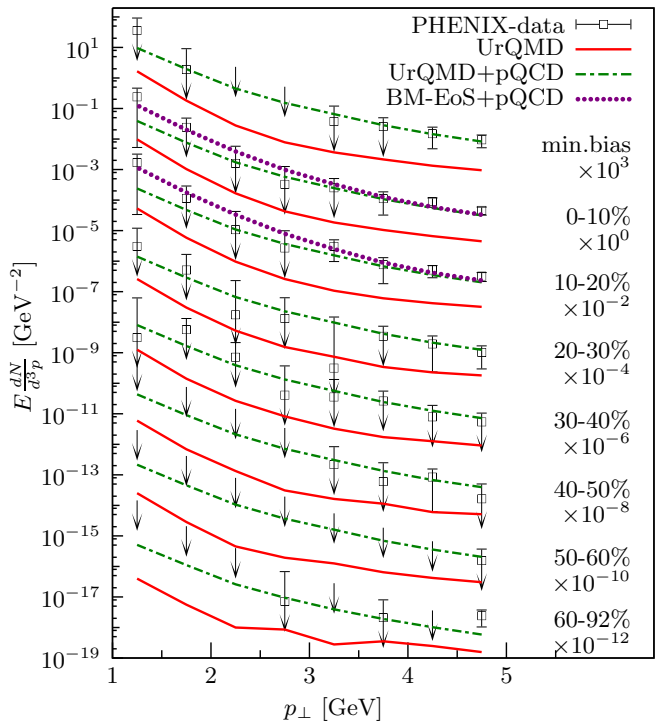}
 \caption{(Color Online) Comparison of the data from the
 PHENIX-collaboration~\cite{nucl-ex/0503003} (black squares) to cascade
 calculations (red solid lines) for central to peripheral collisions. The
 green dash-dotted lines show the sum of
 pQCD-calculations~\cite{Gordon:1993qc,nucl-ex/0503003} and the cascade
 contribution. For the most central collisions, 00-10\% and 10-20\%,
 the spectra from hybrid calculations with the BM-EoS plus pQCD-contribution
 are shown (violett dotted lines).
 }
 \label{fig:auau:200:ca}
\end{figure}
\begin{table}
 \begin{tabular}{l|r@{.}l@{$\pm$}r@{.}l|r@{.}l@{$\pm$}r@{.}l|r@{.}l}
  Centrality & \multicolumn{4}{c|}{$T_{\rm slope}~[$MeV$]$} & \multicolumn{4}{c|}{$A~[$GeV$^{-2}]$} & \multicolumn{2}{c}{$\chi^2$/d.o.f.} \\ \hline
  00\%-10\% & 231&9 & 9&4 & 2&39 & 0&67 & 0&038 \\
  00\%-92\% & 231&4 & 8&5 & 0&41 & 0&11 & 0&032 \\
  10\%-20\% & 234&0 &10&0 & 1&26 & 0&37 & 0&041 \\
  20\%-30\% & 239&0 &11&4 & 0&56 & 0&18 & 0&049 \\
  30\%-40\% & 239&0 &13&1 & 0&27 & 0&10 & 0&065 \\
  40\%-50\% & 243&0 &13&4 & 0&12 & 0&04 & 0&064 \\
  50\%-60\% & 235&4 & 8&8 & (5&64 & 1&43)$\cdot 10^{-2}$ & 0&032 \\
  60\%-92\% & 250&5 &11&8 & (6&91 & 2&08)$\cdot 10^{-3}$ & 0&044 \\
 \end{tabular}
 \caption{Fit results for the low-$p_\bot$-part ($p_\bot < 2.5$~GeV) of the
 cascade calculations of Au+Au-collisions at $\sqrt{s_{\sf NN}} = 200$~GeV
 (see Fig.~\ref{fig:auau:200:ca}). The fit function is $f(p_\bot) = A
 \exp\left (-\frac{p_\bot}{T_{\rm slope}}\right )$.}
 \label{tab:fit:cascade}
\end{table}

The comparison between direct photon spectra at low and intermediate
transverse momentum $p_\bot$ from cascade calculations and data from the
PHENIX collaboration~\cite{nucl-ex/0503003} for Au+Au-collisions at
$\sqrt{s_{\sf NN}} = 200$~GeV is shown in Fig.~\ref{fig:auau:200:ca}.
One clearly observes that the hadronic transport model (full lines) does not
saturate the upper limits of the experimental data. In
all centrality bins, the prompt photon yield is significantly larger
than predicted by the hadronic cascade. The ratio between pQCD and hadronic
contributions is fairly constant among the centrality bins. For comparison,
Fig.~\ref{fig:auau:200:ca} also shows the spectra obtained with the hybrid
model using the Bag Model EoS (BM-EoS) for the two most central bins,
00-10\% and 10-20\%, which agrees nicely with the data. Thermal fits to the
low-$p_\bot$-parts of the cascade spectra show inverse slope parameters of $T_{\sf
slope} \approx 235$~MeV throughout the centrality bins, see
Table~\ref{tab:fit:cascade}.

\begin{figure}
 \input{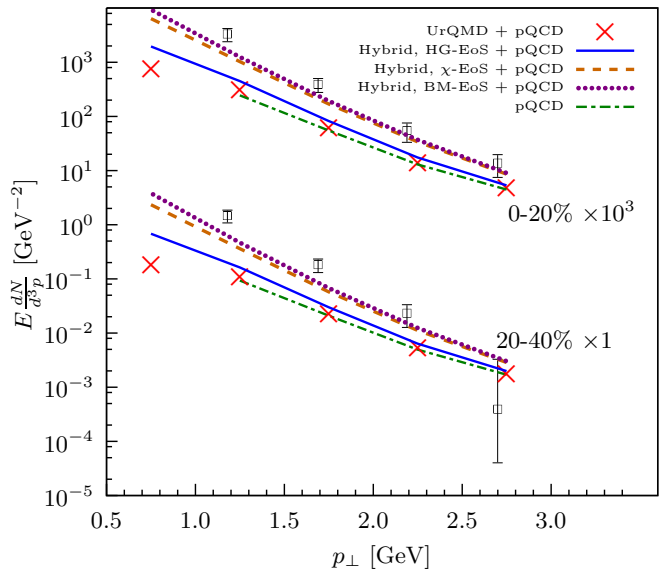}
 \caption{(Color Online) Comparison of data from the
 PHENIX-collaboration~\cite{arXiv:0804.4168} (black squares) to cascade
 caculations (red crosses) and hybrid-model calculations with HG-EoS (blue
 solid lines), $\chi$-EoS (orange dashed lines) and BM-EoS (violett dotted
 lines) for central (0-20~\%) and mid-central (20-40~\%) collisions. The
 contribution from initial
 pQCD-scatterings~\cite{Gordon:1993qc,arXiv:0804.4168} have been added to
 all spectra. The spectra from central collisions have been scaled by a
 factor of $10^3$ to enhance readability.
 }
 \label{fig:auau:200}
\end{figure}

A more detailed exploration of the low-$p_\bot$-part of the direct photon
calculation is shown in Fig.~\ref{fig:auau:200}. Here, the low-$p_\bot$-data
obtained by extrapolating the dilepton yield to zero invariant
mass~\cite{arXiv:0804.4168} for central (00-20\%) and mid-central (20-40\%)
is shown in comparison to cascade calculations (red crosses) and hybrid
calculations with hadron gas EoS (HG-EoS, solid blue lines), chiral EoS
($\chi$-EoS, dashed orange lines) and bag model EoS (BM-EoS, dotted violett
lines) and prompt (pQCD) photon calculations. All calculated spectra include the
$\langle N_{\sf coll} \rangle$-scaled prompt photon contribution.

In both centrality-bins, the direct photon spectra obtained with the BM-EoS
and $\chi$-EoS, which include a phase transition to a deconfined state of
matter, are significantly higher than the hadronic HG-EoS-calculations and
agree with the measured data.

\begin{figure}
 \input{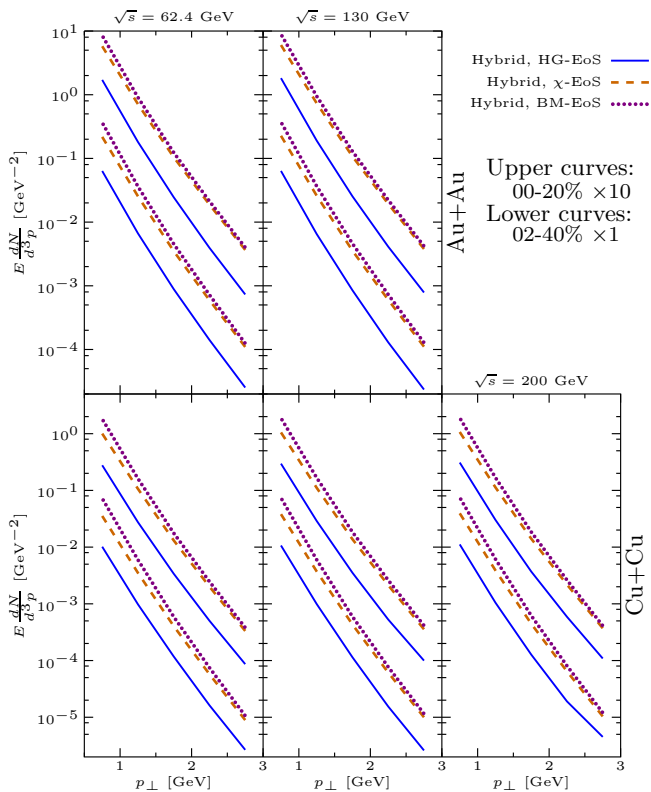}
 \caption{(Color Online). Direct photon spectra calculated with the Hybrid
 model and HG-EoS (solid blue lines), $\chi$-EoS (dashed orange lines) and
 BM-EoS (dotted violett lines) without prompt photon contribution. The left
 panels show calculations for $\sqrt{s_{\sf NN}} = 62.4$~GeV, the middle
 panels show calculations for $\sqrt{s_{\sf NN}} = 130$~GeV and the right
 panel shows calculations for $\sqrt{s_{\sf NN}} = 200$~GeV. The upper
 panels show calculations for Au+Au-collisions, while the lower panel show
 calculations for Cu+Cu-collisions. In each panel, the upper curves are
 central collisions (00-20\%) and the lower curves are mid-central
 collisions (20-40\%). }
 \label{fig:panels}
\end{figure}

A similar picture presents itself in Au+Au-collisions at lower incident
energy $\sqrt{s_{\sf NN}} = 62.4$~GeV and $\sqrt{s_{\sf NN}} = 130$~GeV,
shown in the upper panels of Figure~\ref{fig:panels}. The cascade
calulations have been omitted from the Figure for clarity. 

Thermal fits to the spectra (see Table~\ref{tab:fit:auau}) show inverse
slope parameters in the range from $233 < T_{\sf slope} < 262$~MeV, with the
cascade calculations showing the smallest and the $\chi$-EoS hybrid
calculations showing the largest values of $T_{\sf slope}$. HG-EoS and
BM-EoS calculations show similar inverse slope parameters. The integrated
yield $A$ is highest in BM-EoS hybrid calculations. The spectra from the
hybrid calculations are rather similar for the different beam energies.

\begin{table}
 \begin{tabular}{l|l|r|r@{.}l@{$\pm$}r@{.}l|r@{.}l@{$\pm$}r@{.}l|r@{.}l}
  $\sqrt{s_{\sf NN}}$ & EoS & centr. & \multicolumn{4}{c|}{$T_{\sf
  slope}~[$MeV$]$} & \multicolumn{4}{c|}{$A~[$GeV$^{-2}]$} &
  \multicolumn{2}{c}{$\frac{\chi^2}{\sf d.o.f.}$} \\ \hline
  200 & \color{red}    Transport  &  0-20\% & 232&5 & 9&8 & 1&65 & 0&48 & 0&041 \\
  200 & \color{blue}   HG-EoS     &  0-20\% & 246&7 & 8&6 & 3&63 & 0&83 & 0&025 \\
  200 & \color{orange} $\chi$-EoS &  0-20\% & 261&9 & 8&7 &10&13 & 2&05 & 0&020 \\
  200 & \color{violett}BM-EoS     &  0-20\% & 251&4 & 9&7 &16&37 & 4&03 & 0&029 \\
  200 & \color{red}    Transport  & 20-40\% & 237&3 &12&1 & 0&38 & 0&13 & 0&057 \\
  200 & \color{blue}   HG-EoS     & 20-40\% & 243&4 & 8&3 & 1&32 & 0&30 & 0&025 \\
  200 & \color{orange} $\chi$-EoS & 20-40\% & 253&0 & 8&0 & 4&11 & 0&82 & 0&020 \\
  200 & \color{violett}BM-EoS     & 20-40\% & 240&6 & 9&0 & 7&61 & 1&90 & 0&030 \\\hline
  130 & \color{red}    Transport  &  0-20\% & 232&5 & 9&1 &(9&87 & 2&67)$^\ast$ & 0&035 \\
  130 & \color{blue}   HG-EoS     &  0-20\% & 246&3 & 8&5 & 3&42 & 0&66 & 0&024 \\
  130 & \color{orange} $\chi$-EoS &  0-20\% & 261&2 & 8&5 & 9&67 & 1&93 & 0&019 \\
  130 & \color{violett}BM-EoS     &  0-20\% & 250&2 & 9&6 &15&84 & 3&88 & 0&039 \\
  130 & \color{red}    Transport  & 20-40\% & 257&2 &11&3 &(5&48 & 1&50)$^+$ & 0&036 \\
  130 & \color{blue}   HG-EoS     & 20-40\% & 242&4 & 7&6 & 1&26 & 0&26 & 0&021 \\
  130 & \color{orange} $\chi$-EoS & 20-40\% & 252&7 & 7&9 & 4&01 & 0&80 & 0&019 \\
  130 & \color{violett}BM-EoS     & 20-40\% & 240&6 & 8&8 & 7&46 & 1&82 & 0&029 \\ \hline
  62.4& \color{red}    Transport  &  0-20\% & 242&1 &13&5 &(5&29 & 1&95)$^\ast$ & 0&066 \\
  62.4& \color{blue}   HG-EoS     &  0-20\% & 247&3 & 8&1 & 3&19 & 0&67 & 0&022 \\
  62.4& \color{orange} $\chi$-EoS &  0-20\% & 261&8 & 8&2 & 9&24 & 1&78 & 0&018 \\
  62.4& \color{violett}BM-EoS     &  0-20\% & 250&3 & 9&5 &15&13 & 3&65 & 0&028 \\
  62.4& \color{red}    Transport  & 20-40\% & 232&8 & 9&4 &(4&18 & 1&16)$^+$ & 0&038 \\
  62.4& \color{blue}   HG-EoS     & 20-40\% & 245&8 & 8&0 & 1&21 & 0&26 & 0&022 \\
  62.4& \color{orange} $\chi$-EoS & 20-40\% & 253&9 & 7&7 & 3&82 & 0&73 & 0&018 \\
  62.4& \color{violett}BM-EoS     & 20-40\% & 240&8 & 8&6 & 7&33 & 1&74 & 0&028 \\ \hline
 \end{tabular}\\
 {\footnotesize $^\ast$: $\times 10^{-2}$,\quad $^+$: $\times 10^{-3}$}
 \caption{Fit results for the low-$p_\bot$-part ($p_\bot < 2.5$~GeV) of the
 spectra from central (0-20\%) and mid-central (20-40\%) Au+Au-collisions.
 The fit function is $f(p_\bot) = A \exp\left (-\frac{p_\bot}{T_{\rm
 slope}}\right )$. The data are shown in Figure~\ref{fig:auau:200} (for
 $\sqrt{s_{\sf NN}} = 200$~GeV) and Figure~\ref{fig:panels} ($\sqrt{s_{\sf
 NN}} = 62.4$~GeV, upper left panel and $\sqrt{s_{\sf NN}} = 130$~GeV, upper
 central panel). }
 \label{tab:fit:auau}
\end{table}

Hybrid model calculations for central (0-20\%) and mid-central (20-40\%)
Cu+Cu-collisions are shown in the lower panels of Figure~\ref{fig:panels}
for all EoS. The thermal fits (see Table~\ref{tab:fit:cucu}) again show no
significant energy dependence of inverse slope parameter $T_{\sf slope}$ or
yield $A$. We observe a clear ordering of the total yield between the
Equations of State, with yield from the BM-EoS calculations being higher
than that of the $\chi$-EoS, and both yields exceeding that of HG-EoS
calculations. However, the inverse slope parameters are similar in HG-EoS
and $\chi$-EoS calculations but significantly lower in BM-EoS calculations.

\begin{table}
 \begin{tabular}{l|l|c|r@{.}l@{$\pm$}r@{.}l|r@{.}l@{$\pm$}r@{.}l|r@{.}l}
  $\sqrt{s_{\sf NN}}$ & EoS & centr. & \multicolumn{4}{c|}{$T_{\sf
  slope}~[$MeV$]$} & \multicolumn{4}{c|}{$A~[$GeV$^{-2}]$} &
  \multicolumn{2}{c}{$\frac{\chi^2}{\sf d.o.f.}$} \\ \hline
  200 & \color{blue}   HG-EoS    &  0-20\% & 252&0 & 9&6 &(4&84 & 1&38)$^\times$ & 0&057 \\
  200 & \color{orange} $\chi$-EoS&  0-20\% & 251&5 & 7&3 & 1&77 & 0&39 & 0&033 \\
  200 & \color{violett}BM-EoS    &  0-20\% & 237&7 & 7&8 & 3&61 & 0&94 & 0&047 \\
  200 & \color{blue}   HG-EoS    & 20-40\% & 254&6 &13&2 &(1&61 & 0&62)$^\times$ & 0&103 \\
  200 & \color{orange} $\chi$-EoS& 20-40\% & 242&9 & 7&0 &(7&25 & 1&63)$^\times$ & 0&036 \\
  200 & \color{violett}BM-EoS    & 20-40\% & 229&2 & 7&5 & 1&60 & 0&43 & 0&051 \\\hline
  130 & \color{blue}   HG-EoS    &  0-20\% & 250&0 & 9&3 &(4&78 & 1&35)$^\times$ & 0&056 \\
  130 & \color{orange} $\chi$-EoS&  0-20\% & 250&9 & 7&1 & 1&76 & 0&37 & 0&031 \\
  130 & \color{violett}BM-EoS    &  0-20\% & 238&1 & 7&9 & 3&56 & 0&93 & 0&048 \\
  130 & \color{blue}   HG-EoS    & 20-40\% & 240&4 & 7&7 &(1&99 & 0&50)$^\times$ & 0&044 \\
  130 & \color{orange} $\chi$-EoS& 20-40\% & 242&8 & 7&1 &(6&99 & 1&59)$^\times$ & 0&036 \\
  130 & \color{violett}BM-EoS    & 20-40\% & 228&5 & 7&7 & 1&58 & 0&44 & 0&054 \\\hline
  62.4& \color{blue}   HG-EoS    &  0-20\% & 248&2 & 7&7 &(4&71 & 1&11)$^\times$ & 0&039 \\
  62.4& \color{orange} $\chi$-EoS&  0-20\% & 250&2 & 6&8 & 1&71 & 0&35 & 0&029 \\
  62.4& \color{violett}BM-EoS    &  0-20\% & 236&8 & 7&4 & 3&52 & 0&88 & 0&044 \\
  62.4& \color{blue}   HG-EoS    & 20-40\% & 242&8 & 6&9 &(1&87 & 0&41)$^\times$ & 0&034 \\
  62.4& \color{orange} $\chi$-EoS& 20-40\% & 241&7 & 6&3 &(6&71 & 1&37)$^\times$ & 0&029 \\
  62.4& \color{violett}BM-EoS    & 20-40\% & 227&0 & 6&7 & 1&62 & 0&40 & 0&042 \\ \hline
 \end{tabular}\\
 {\footnotesize $^\times$: $\times 10^{-1}$}
 \caption{Fit results for the low-$p_\bot$-part ($p_\bot < 2.5$~GeV) of the
 spectra from central (0-20\%) and mid-central (20-40\%) Cu+Cu-collisions.
 The fit function is $f(p_\bot) = A \exp\left (-\frac{p_\bot}{T_{\rm
 slope}}\right )$. The data are shown in Figure~\ref{fig:panels}, lower
 panels. }
 \label{tab:fit:cucu}
\end{table}

\section{Summary}\label{sec:summary}

We examined the direct photon spectra obtained with a transport and a
transport+hydrodynamics hybrid model for collisions of Au+Au and Cu+Cu at
energies of $\sqrt{s_{\sf NN}}$ = 62.4, 130 and 200 GeV. We find that the
hadronic models (transport model and hybrid model with Hadron Gas EoS)
underpredict the data, while calculations with a deconfined state of matter
(hybrid model with Chiral or Bag Model EoS) fit the data much better.

Thermal fits to the data show no significant beam energy dependence on the
spectra. The inverse slope parameters obtained by fitting the low-transverse
momentum part of the spectra are in the range of $227 < T_{\sf slope} <
262$~MeV, which is significantly above the expected transition temperature
to deconfined matter.

Prompt photons from the initial early hard proton-proton scatterings are
found to be a significant source of direct photon emission above $p_\bot =
3.5$~GeV, if an EoS with phase transition is assumed, and is dominant
throughout all $p_\bot$ if a purley hadronic scenario is assumed.

\section{Outlook}

Future work with this model will include the extraction of radial and
elliptic flow parameters $v_1$ and $v_2$ for more differential analyses.
Also, the influence of changing the criteria for the transition between the
transport- and hydrodynamic phases in the hybrid model will be examined in
the future.

\section{Acknowledgements}

This work has been supported by the Frankfurt Center for Scientific
Computing (CSC), the GSI and the BMBF. B.\ B\"auchle gratefully acknowledges
support from the Deutsche Telekom Stiftung, the Helmholtz Research School on
Quark Matter Studies and the Helmholtz Graduate School for Hadron and Ion
Research. This work was supported by the Hessian LOEWE initiative through
the Helmholtz International Center for FAIR.

The authors thank Elvira Santini for valuable discussions and Henner
B\"usching for experimental clarifications.

%\appendix

\end{document}